# Quantum interference in spontaneous emission from a V-type three-level atom in photonic crystals

Chih-Hsien Huang,[1,2] Jing-Nuo Wu,[3] Yen-Yin Li,[2] Szu-Cheng Cheng,[3,*] and Wen-Feng Hsieh[1,2,*]

[1]Institute of Electro-Optical Science and Engineering, National Cheng Kung University, 1 Dahsueh Rd., Tainan 701, Taiwan

[2]Department of Photonics and Institute of Electro-Optical Engineering, National Chiao Tung University, 1001 Tahsueh Rd., Hsinchu 300, Taiwan

[3]Department of Physics, Chinese Culture University, Yangming Mt., Taipei 111, Taiwan

**Abstract**

Studying the spontaneous emission of a V-type three-level atom embedded in a photonic crystal (PC) by fractional calculus, we found that the atomic excited states in the anisotropic PC can be expressed as a superposition of four dressed states analytically. Through detuning two allowed atomic transition energies with respect to the photonic band edge, the coupling between these two transitions leads to three dynamic regimes, namely non-Markovian decay, damped quantum interference and quantum interference, classified by the numbers of contributed bounded dressed states. From the degree of quantum interference of two atomic transitions, we found the energy exchange between the atom and PC reservoir is the lowest as the excited states become degenerate but with maximum quantum interference when the atom is prepared at one of the excited states. The results also show that excited states prefer to stay out of phase at all detuning energy except for near degenerate. Therefore, we can control the spontaneous emission rate not only by the amount of detuning frequencies but also the relative phase of initial states.

[*] to whom correspondence should be addressed.
Email: sccheng@faculty.pccu.edu.tw and wfhsieh@mail.nctu.edu.tw.

# Ⅰ. INTRODUCTION

Photonic crystals (PCs) are periodic dielectric structures [1] providing a photonic band gap (PBG) and redistribution of photon density of state (DOS) to control the spontaneous emission (SE) [2-8]. Within the PBG, light is forbidden to propagate in a PC so that the SE is depressed. Near the PBG, however, the anomalous photon DOS makes the Markovian approximation of SE as in free space invalid. The non-Markovian photon-atom interaction gives rise to rapid multi-atom switch with low quantum noise and laser-like collective atomic emission [9-10]. It also offers the key technology for manipulating light such as light emitting devices [11] and solar cells [12].

The SE in three-level atom systems, including V-type [13-15], cascade-type [16] and Λ-type arrangements [4-5, 17-18], are of particular interest due to the quantum interference between two allowed transitions. The quantum interference between different atomic transitions or atomic coherence in the V-type system can lead to population trapping, phase-sensitive amplification, and laser without inversion. [14, 19-20]

The study of the V-type atom systems embedded in PCs was focused mainly on their emission spectra [15, 21-22], but rarely on the evolution of excited-state population due to the requirement for dealing with complex integration in the inverse Laplace transformation [14, 23], which has encountered the multi-value problem. In order to simplify the computation, the band edge of the PC had been usually assumed midway between two excited levels [24] in the isotropic models [23]. However, the structures of PCs are quite anisotropic in general and the band edge exactly localize between two excited levels are far from a generalized case. Besides, the analytic expression of excited-state population is hard to be obtained with complex integral, and the results of the anisotropic model even conflict with experiment results [25] in the system of a two-level atom [6]. Therefore, a more precise approach using the fractional calculus [26] is needed to deal with the quantum interference of a three-level atom embedded in a PC analytically.

In this paper, we first use the fractional calculus to derive the governed equations describing the time evolution of excited states in a V-type atom embedded in a PC. The results reveal that the population of excited states is a linear combination of four dressed states (DSs) in the anisotropic model. Second, by analyzing the properties of four DSs, the dynamic behavior of the exited atoms is discussed. In order to verify the correctness of our derivation, we reduce the system into special and degenerated cases and compare them with the previous researches [4, 6, 8, 14, 23, 27] in the following section. Finally, the quantum interference and correlation between two excited states are discussed.

## II. THE DYNAMICS OF QUANTUM INTERFERENCE

Consider a three-level atom with a ground state $|b\rangle$ and two excited states $|a_1\rangle$ and $|a_2\rangle$, shown in Fig. 1, is embedded inside a PC. The allowed transitions from two excited states $|a_1\rangle$ and $|a_2\rangle$ to the ground state $|b\rangle$ have resonant frequencies $\omega_1$ and $\omega_2$, respectively, located around the PC band edge. Therefore, the total Hamiltonian of the system with atom-field interaction is $\hat{H} = \hat{H}_0 + \hat{H}_I$ with the non-interaction Hamiltonian,

$$\hat{H}_0 = \hbar\omega_1\sigma_{a_1a_1} + \hbar\omega_2\sigma_{a_2a_2} + \sum_{\mathbf{k}}\hbar\omega_{\mathbf{k}}a_{\mathbf{k}}^+ a_{\mathbf{k}} \tag{1}$$

and the interaction Hamiltonian,

$$\hat{H}_I = i\hbar\left[\sum_{\mathbf{k}}\left(g_{1\mathbf{k}}a_{\mathbf{k}}^+\sigma_{ba_1} + g_{2\mathbf{k}}a_{\mathbf{k}}^+\sigma_{ba_2} - g_{1\mathbf{k}}a_{\mathbf{k}}\sigma_{a_1b} - g_{2\mathbf{k}}a_{\mathbf{k}}\sigma_{a_2b}\right)\right]. \tag{2}$$

Here, $\sigma_{ij} = |i\rangle\langle j|$, ($i, j = a_1, a_2$ or $b$), $\omega_{\mathbf{k}}$, $a_{\mathbf{k}}^+$ and $a_{\mathbf{k}}$ are the radiative frequency, the creation and annihilation operators of mode $\mathbf{k}$ of the reservoir; the atom-field coupling coefficients of two excited states, $g_{1\mathbf{k}}$ and $g_{2\mathbf{k}}$ are [28]

$$g_{n\mathbf{k}} = \frac{\omega_n d_n}{\hbar}\left(\frac{\hbar}{2\varepsilon_0\omega_{\mathbf{k}}V}\right)^{1/2}\mathbf{e}_{\mathbf{k}}\cdot\mathbf{u}_d. \tag{3}$$

We have assumed a fixed atomic dipole moment $\mathbf{d}_n = d_n\mathbf{u}_d$ which is independent of position with quantization volume $V$ and wave propagation direction $\mathbf{e}_\mathbf{k}$. The dispersion relation of a PC near the band edge $\omega_c$ can be approximately expressed as $\omega_k = \omega_c + D(k - k_c)^2$ for the isotropic model and $\omega_k = \omega_c + D(\vec{k} - \vec{k}_c)^2$ for the anisotropic model, where $D \simeq f_\mathbf{k}\frac{\omega_c}{k_c^2}$ is the curvature near $\omega_c$ with scaling factor $f_\mathbf{k}$ for different $\mathbf{k}$ directions.

Because 3D PCs are highly anisotropic structures, in the following discussion we shall mainly focus on the anisotropic model although some results of the isotropic model will also be derived for comparison. We let two dipole vectors parallel to each other because the interference are similar with parallel and antiparallel dipoles, and there is only the combination of two systems with no interaction between these two transitions if these two dipoles are orthogonal [14]. Having $\omega_1 - \omega_c = \Delta_1$ and $\omega_2 - \omega_c = \Delta_2$ with $|\Delta_1|$, $|\Delta_2| \ll \omega_c$, we further assume $g_{1\mathbf{k}} = g_{2\mathbf{k}} = g_\mathbf{k}$. The state vector $|\psi(t)\rangle$ at a time instant $t$ thus can be expressed as the superposition of bare states $|a_{1,2}, 0\rangle$ and $|b, 1\rangle$ of the one-photon sector,

$$|\psi(t)\rangle = A_1(t)e^{-i\omega_1 t}|a_1, 0\rangle + A_2(t)e^{-i\omega_2 t}|a_2, 0\rangle + \sum_\mathbf{k} B_\mathbf{k}(t)e^{-i\omega_\mathbf{k} t}|b, 1_\mathbf{k}\rangle. \quad (4)$$

Here $|a_{1,2}, 0\rangle$ describes the atom in its excited state $|a_1\rangle$ or $|a_2\rangle$ with no photons present and $|b, 1_\mathbf{k}\rangle$ represents for the atom in its ground state $|b\rangle$ and a single photon in mode $\mathbf{k}$.

From the time-dependent Schrödinger equation, we obtain

$$A'_n(t) = -\sum_\mathbf{k} g_\mathbf{k} B_\mathbf{k}(t) e^{-i(\omega_\mathbf{k} - \omega_n)t}; \quad (5)$$

$$B'_\mathbf{k}(t) = g_\mathbf{k}[A_1(t)e^{-i(\omega_1 - \omega_\mathbf{k})t} + A_2(t)e^{-i(\omega_2 - \omega_\mathbf{k})t}] \quad (6)$$

with $n = 1$ or 2. Integrating Eq. (6), we obtain

$$B_k(t) = g_\mathbf{k} \int_0^t [A_1(\tau)e^{-i(\omega_1 - \omega_\mathbf{k})\tau} + A_2(\tau)e^{-i(\omega_2 - \omega_\mathbf{k})\tau}] d\tau. \quad (7)$$

Substituting Eq. (7) into Eq. (5), we have

$$\vec{A}'(t) = -\int_0^t \vec{G}(t - \tau) \cdot \vec{A}(\tau) d\tau, \quad (8)$$

where $\vec{G}(t-t')$ is the memory kernel given by

$$\vec{G}(t - \tau) = \sum_\mathbf{k} |g_\mathbf{k}|^2 \begin{bmatrix} e^{-i(\omega_\mathbf{k} - \omega_1)(t-\tau)} & e^{-i\omega_\mathbf{k}(t-\tau) + i\omega_1 t - i\omega_2 \tau} \\ e^{-i\omega_\mathbf{k}(t-\tau) + i\omega_2 t - i\omega_1 \tau} & e^{-i(\omega_\mathbf{k} - \omega_2)(t-\tau)} \end{bmatrix} \quad (9)$$

and $\vec{A}(\tau) = \begin{bmatrix} A_1(\tau) \\ A_2(\tau) \end{bmatrix}$.

In the continuum limit, one can replace $\sum_{\mathbf{k}}$ by $V \int_0^\infty \rho(\omega) d\omega$ [10] so that the memory Kernel is

$$\vec{G}(t-\tau) = \frac{r_d e^{i\frac{5\pi}{4}}}{2\sqrt{\pi}(t-\tau)^{3/2}} \begin{bmatrix} e^{i\Delta_1(t-\tau)} & e^{i\Delta_1 t - i\Delta_2 \tau} \\ e^{-i\Delta_1 t + i\Delta_2 \tau} & e^{i\Delta_2(t-\tau)} \end{bmatrix}, \qquad (10)$$

where $r_d = \omega^2 d^2 / (8\hbar\varepsilon_0 \omega_c D^{3/2} \pi)$ is the coupling constant. Here we have used the photon DOS of the anisotropic model, $\rho(\omega) = \sqrt{\omega - \omega_c} \Theta(\omega - \omega_c)/(4\pi^2 D^{3/2})$ with $\Theta(u)$ being the Heaviside step function.

By substituting Eq. (10) into Eq. (8) and making a transformation of $A_1(t) = e^{i\Delta_1 t} C_1(t)$ and $A_2(t) = e^{i\Delta_2 t} C_2(t)$, we have

$$\frac{d}{dt} C_n(t) + i\Delta_n C_n(t) = -\frac{(r_d e^{i5\pi/4})}{2\sqrt{\pi}} \int_0^t \frac{C_1(\tau) + C_2(\tau)}{(t-\tau)^{3/2}} d\tau. \qquad (11)$$

From the definition of Riemann-Liouville fractional differentiation operator, [26]

$$\frac{d^\alpha}{dt^\alpha} u(t) = \frac{1}{\Gamma(-\alpha)} \int_0^t (t-s)^{-\alpha-1} u(s) ds, \qquad (12)$$

Eq. (11) can be expressed as

$$\frac{d}{dt} C_n(t) + i\Delta_n C_n(t) = -r_d e^{i\pi/4} \frac{d^{-1/2}}{dt^{-1/2}} [C_1(t) + C_2(t)], \qquad (13)$$

where $\Gamma(x)$ is the gamma function. We further apply the integral operator $d^{-1}/dt^{-1}$ followed by the differential operator $d^{1/2}/dt^{1/2}$ to Eq. (13) to get the fractional quantum Langevian equation,

$$\frac{d^{1/2}}{dt^{1/2}}C_n(t)+i\Delta_n\frac{d^{-1/2}}{dt^{-1/2}}C_n(t)+r_d e^{i\pi/4}[C_1(t)+C_2(t)]=\frac{C_n(0)}{\sqrt{\pi}}t^{-1/2}. \quad (14)$$

Taking Laplace transform of these two fractional Langevian equations for $n = 1$ and 2, we have

$$A_n(s+i\Delta_n)=\frac{A_n(0)(s+i\Delta_{3-n})+(-1)^n r_d\sqrt{s}e^{i\frac{\pi}{4}}[A_1(0)-A_2(0)]}{s^2+2r_d e^{i\frac{1}{4}\pi}\sqrt{s}^3+is(\Delta_1+\Delta_2)+ir_d e^{i\frac{1}{4}\pi}\sqrt{s}(\Delta_1+\Delta_2)-\Delta_1\Delta_2}. \quad (15).$$

Let $X = s^{1/2}$, we can then rewrite Eq. (15) as a sum of partial fractions

$$A_n(X+i\Delta_n)=\sum_{m=1}^{4}\frac{{}^n\alpha_m}{(X-X_m)}. \quad (16)$$

Note that the parameters $X_m$ ($m = 1, 2, 3, 4$) of Eq. (16) are the roots of

$$X^4+2r_d e^{i\frac{1}{4}\pi}X^3+i(\Delta_1+\Delta_2)X^2+r_d e^{i\frac{3}{4}\pi}(\Delta_1+\Delta_2)X-\Delta_1\Delta_2=0, \quad (17)$$

and the coefficients ${}^n\alpha_m$ are given by

$$^n\alpha_m=\frac{A_n(0)[X_m^2+i\Delta_{3-n}]-r_d e^{i\frac{\pi}{4}}X_m[A_1(0)-A_2(0)](-1)^n}{\prod_{j=1(\neq m)}^{4}(X_m-X_j)}. \quad (18)$$

From the formula of the inverse fractional Laplace transformation [26]

$$\mathcal{L}^{-1}\left\{\frac{1}{s^{1/2}-X}\right\}=E_t(-\frac{1}{2},X^2)+Xe^{X^2t}, \quad (19)$$

we can get the probability amplitudes of two excited states

$$A_1(t)=e^{it\Delta_1}\sum_{m=1}^{4}{}^1\alpha_m\left[E_t(-\frac{1}{2},X_m^2)+X_m e^{X_m^2 t}\right]=e^{it\Delta_1}\sum_{m=1}^{4}{}^1\alpha_m\left[X_m^2 E_t(\frac{1}{2},X_m^2)+X_m e^{X_m^2 t}\right] \quad (20)$$

and

$$A_2(t) = e^{it\Delta_2} \sum_{m=1}^{4} {}^2\alpha_m \left[ E_t(-\frac{1}{2}, X_m^2) + X_m e^{X_m^2 t} \right], \quad (21)$$

where $E_t(\gamma, a)$ is the fractional exponential function of order $\gamma$ and is defined as

$$E_t(\gamma, a) = t^\gamma \sum_{n=0}^{\infty} \frac{(at)^n}{\Gamma(\gamma+n+1)}. \quad (22)$$

In the isotropic model with the dispersion relation $\omega_k = \omega_c + D(k-k_c)^2$,

$$A_n(s+i\Delta_n) = \frac{A_1(0)(s+i\Delta_{3-n}) + (-1)^n[A_2(0) - A_1(0)]\frac{r_c e^{-i\frac{\pi}{4}}}{\sqrt{s}}}{s^2 + is(\Delta_1 + \Delta_2) + i\frac{r_c e^{-i\frac{\pi}{4}}}{\sqrt{s}}(\Delta_1 + \Delta_2) + 2r_c e^{-i\frac{\pi}{4}}\sqrt{s} - \Delta_1 \Delta_2}, \quad (23)$$

where $r_c = \omega^{7/2} d^2 / (6\hbar\varepsilon_0 \pi c^3)$ with $D \cong \omega_c / k_c^2$. By similar method using the fractional calculus, we can also derive the amplitudes of excited states $|a_1\rangle$ and $|a_2\rangle$ are the linearly combination of five DSs characterized by five $X_m's$, which satisfy

$$X^5 + iX^3(\Delta_1 + \Delta_2) + 2r_c e^{-i\frac{\pi}{4}} X^2 - \Delta_1 \Delta_2 X + ir_c e^{-i\frac{\pi}{4}}(\Delta_1 + \Delta_2) = 0. \quad (24)$$

## Ⅲ. THE PROPERTIES OF DRESSED STATES AND SPONTANEOUS EMISSION

From the previous section, the population amplitudes of the two excited states in an anisotropic PC contributed from four DSs can be expressed as analytical forms such as Eqs. (20) and (21). These two equations can be further written as [6, 26]

$$A_1(t) = e^{it\Delta_1} \sum_{m=1}^{4} {}^1\alpha_m \left[ Y_m Erf(\sqrt{X_m^2 t}) + X_m \right] e^{X_m^2 t}, \quad (25)$$

$$A_2(t) = e^{it\Delta_2} \sum_{m=1}^{4} {}^2\alpha_m \left[ Y_m Erf(\sqrt{X_m^2 t}) + X_m \right] e^{X_m^2 t}, \qquad (26)$$

with $Y_m = \sqrt{X_m^2}$ and $E_t(1/2, X) = Exp(Xt) Erf(\sqrt{Xt})/\sqrt{X}$. Here $Erf(t)$ is the error function. Substituting Eqs. (25) and (26) into Eq. (4), it is obvious that the wavefunction possesses four DS with the frequencies equals to $\omega_c - Im(X_m^2)$, where *Im* represents for the imaginary part. When $X_m^2$ is a complex number, the population of the excited-state contributed from the DS characterized by $X_m$ will behave as decaying. On the other hand, if $X_m^2$ is a pure positive imaginary number, the population contributed from the DS of $X_m$ may oscillate initially due to the error function term and then the oscillation decreases as time passes due to $Erf(\sqrt{X_m^2 t}) = 1$ as $t \to \infty$. Under this circumstance, the amplitude of atomic excited state contributed from the *m*-th dressed state of $X_m$ equals to $2\alpha_m X_m Exp(X_m^2 t)$ at $t = \infty$ as $X_m$ locates in the 1$^{st}$ quadrant, but it equals to 0 as $X_m$ locates in the 3$^{rd}$ quadrant. Therefore, only the DS with $X_m$ having amplitude $\alpha_m Exp(i\pi/4)$ ($\alpha_m > 0$) contributes a bound state to the $|a_1, 0\rangle$ or $|a_2, 0\rangle$ state, otherwise, the DSs are the decaying states. The magnitude of the amplitude of the excited state $|a_1, 0\rangle$ or $|a_2, 0\rangle$ contributed from the *m*-th bound DS is $2\alpha_m X_m$.

Therefore, the time evolution of the excited-state of a three-level atom in an anisotropic PC would behave differently as a result of different numbers of contributed bound DSs and can be categorized into three regimes:

(1) Non-Markovian decay regime with no bound DS. It behaves spontaneous non-Markovian decay with some oscillation initially contributing from either the exponential terms or the interference of 4 unbound DSs.

(2) Damped quantum interference regime with one bounded DS. The excited-state populations will oscillate initially because the strong interaction between photon and the PC reservoir. This oscillation will diminish and finally reach a steady (bound) state.

(3) Quantum interference regime with two bound DSs. The populations will always oscillate due to the interference of two bound DSs. The oscillation or Rabi frequency equals to the frequency difference of these two DSs, i.e., $|Im(X_i^2 - X_j^2)|$.

## IV. POPULATION EVOLUTION OF EXCITED STATES

As mention previously, the quantum interference between two allowed transitions strongly depends on the strength of atom-photon coupling and the detuning of atomic levels respect to the photonic band edge that causes the atomic level splitting or formation of the bound DSs. The dynamic of the atom-photon interaction can be categorized into three regimes, shown in Fig. 2, in terms of the normalized detuning of atomic levels with respect to the coupling strength, i.e., $\Delta_1/r_d$ and $\Delta_2/r_d$, respectively. When both excited states are in the allowed band, i.e., $\Delta_1/r_d$ and $\Delta_2/r_d > 0$, there would be no bounded DS. The electron in the excited states will decay by non-Markovian SE. When one excited state is above the band edge and the other is below the band edge in the gap, one bounded dressed is existed. The dynamics of the excited state behaves as damped quantum interference. When both of the excited states are within the band gap, the two bounded DSs are present to cause the quantum interference except at the degeneration ($\Delta_1 = \Delta_2$).

In the following, we will examine the evolution of excited-state populations corresponding to these three regimes for various initial conditions by setting the initial amplitudes $A_1(0) = cos\theta$ and $A_2(0) = sin\theta$ with $\theta = 0$, $\frac{\pi}{4}$, and $-\frac{\pi}{4}$, respectively.

**A. Non-Markovian decay regime**

When the excited states of the atom are both in the photonic allowed band, the DSs are all unbounded. The atom transfers all its stored energy to the SE propagating field in the PC, therefore, it shows decaying excited-state populations for $\Delta_1/r_d = 0.5$ and $\Delta_2/r_d = 0.25$ in Fig. 3(a). The equally initial-prepared excited atom with θ = π / 4 has the higher decay rate over θ = 0 and θ = - π / 4 cases because there is less energy exchange between two excited states for θ = π / 4. It is worth mentioning that the total decay rate depends on the probabilities of participant decay channels, $^1\alpha_m$ and $^2\alpha_m$, which are determined by initial conditions so that the total decay rate will be slow down as the probability of fast decay channels approach zero. Therefore, the decay rates in Fig. 3(a) show non-Markovian fast decay at the beginning but become slow after evolving a certain time. Such phenomena can also be observed in the carrier relaxation in semiconductor materials [29-30].

**B. Damped quantum interference regime**

As mentioned previously, only one bounded DS exists as one of the excited atom states in a V-type atomic system is in the allowed band, e.g., the state $|a_1\rangle$ with $\Delta_1/r_d$ = 0.5 and the other within the band gap (state $|a_2\rangle$ with $\Delta_2/r_d$ = -0.5). Under this circumstance, the dynamics reveals fast damped quantum interference initially and finally reaches a stationary bound state as shown in Fig. 3(b). Although the state $|a_1\rangle$ is in the allowed band, its population does not completely decay to zero due to the existence of one bounded DS and is much smaller than that of the state $|a_2\rangle$ located within the gap. The initial fast damped quantum interference occurs in the time interval having $r_d t = 0 \sim 25$ is caused by energy transfer between the atom and the PC reservoir through interference of two excited states before relaxing the decaying DSs population to photon.

**C. Quantum interference regime**

As the energies of excited states are both within the band gap, two of the DSs are bounded states. The energy would transfer between these two DSs so the population of two excited states will oscillate periodically after the decaying DSs are diminished as shown in Fig. 3(c) with $\Delta_1/r_d$ = -0.5 and $\Delta_2/r_d$ = -0.25. The Rabi frequency equals to the frequency difference of this bounded DSs, i.e., $|Im(X_i^2 - X_j^2)|$. The bounded

DS energy is related to the energy of the excited state, i.e., the higher energy of an excited state, the higher eigenenergy of a bounded DS is. Therefore, the larger difference of the excited atomic energies corresponds to the higher Rabi frequency.

## V. SPECIAL CASES

### A. Asymptotic two-level atom

When one of the unperturbed atom states, e.g., the state $|a_2\rangle$, locates far above the band edge of the PC, it should hardly interact with the other state and act as in the free space. Therefore, let $\Delta_2$ be equal to infinite in Eq. (17), we can rewrite this equation as

$$X^2 + r_d e^{i\frac{\pi}{4}} X + i\Delta_1 = 0, \tag{27}$$

which corresponds to the characteristic equation of two level system of the anisotropic model [6]. It possesses one bounded DS when $\Delta_1 < 0$ and no bounded DS when $\Delta_1 > 0$. Under this circumstance, the three-level system is asymptotic to a two-level one.

However, in the isotropic model, the DSs are characterized by the solutions [4, 8] of

$$X^3 + i\Delta_1 X + r_c e^{-i\frac{\pi}{4}} = 0 \tag{28}$$

from Eq. (24). It is easy to verify that there is always at least a bounded DS in this equation. However, when the detuning of the excited state $|a_1\rangle$ is farther from the photonic band edge, the smaller contribution of the bounded state is. The population behavior will also mainly dominated by decaying states.

### B. Degenerate excited states

As the detuning of these two excited states are the same or degenerate, $\Delta_1 = \Delta_2 = \Delta$, Eq. (17) becomes

$$(X^2 + i\Delta)(X^2 + 2r_d e^{i\frac{\pi}{4}} X + i\Delta) = 0. \qquad (29)$$

with two roots being $X^2 = i\Delta$ or $X = \pm\sqrt{i\Delta}$; from Eq. (18), the corresponding amplitudes ($^n\alpha_m$) of these two roots are zero if two excited states are equally populated. In this case, four DSs reduce to two and it is equivalent to the system of two-level atom embedded in the PC but with the coupling constant being doubled due to the double degeneracy.

**C. Photonic band edge at the midway of two excited states**

When these two excited states are oppositely detuned from the photonic band edge, namely $\Delta_1 = -\Delta_2 = \Delta$, as discussed in [23] for isotropic model and in [24] of spontaneous emission spectrum for anisotropic model, the characteristic equation for the isotropic model from Eq. (24) is

$$X(X^4 + 2r_c e^{-i\frac{\pi}{4}} X + \Delta^2) = 0. \qquad (30)$$

One of the roots of Eq. (30) is zero, which contributes nothing to the probability amplitudes of the excited states and the other four roots are the same with previous research [23]. The probability amplitudes of these two excited states can also be written as Eqs. (20) and (21) in terms of these four nonzero roots of Eq. (30) which are separately located in four quadrants of the complex plane. As discussed above, only the root located in the 1$^{st}$ quadrant contributes to a bounded DS. Therefore, there would be only one bounded DS in this case and the population of both excited states will be a constant in a long time scale.

The characteristic equation for the anisotropic model from Eq. (17) is

$$X^4 + 2r_d e^{i\frac{\pi}{4}} X^3 + \Delta^2 = 0, \qquad (31)$$

which is the same as the results of [24]. In the this model, there also exists one bounded DS, but the frequency difference between the band edge and bounded DS ($\omega_c - \omega$), which equals to the square of the positive and real solution of Eq. (31) by letting $X = Y\,Exp(i\pi/4)$, is less negative than that in the isotropic model. The higher frequency ($\omega$) DS would have the less stored energy in the atom that does not transfer to photon propagating in the PC if the same coupling constants ($\gamma_d$ and $\gamma_c$) and initial conditions are assumed. It is because the photon DOS near the band edge ($\omega_c$) is less in the anisotropic PC than in the isotropic PC that causes more photons emission to the anisotropic PC due to the smaller photon-atom interaction.

# VI. QUANTUM INTERFERENCE AND CORRELATION OF TWO EXCITED STATES

In order to characterize the energy exchange between two forbidden excited states through coupling with the PC or via quantum interference of two allowed transitions, we define the quantum interference as the degree of energy transfer from the state $|a_1\rangle$ to state $|a_2\rangle$ by the difference of maximum and minimum populations of state $|a_2\rangle$ as $Q_2$ when the decaying DSs are diminished; $Q_T$ is the variation of total excited-state population representing the total energy exchange between the atom and PC. Let the state $|a_2\rangle$ be fixed within the PBG with the detuning $\Delta_2 = -r_d$, we can see in Fig. 4(a), there would have the largest quantum interference between two excited-state transitions around the degenerate ($\Delta_1 = \Delta_2$) where $Q_T = 0$ for $\theta = 0$. It is understandable that in order to have strong interference of two transitions from excited states $|a_1\rangle$ and $|a_2\rangle$ to $|b\rangle$, between which the transition is forbidden and the coupling needs to be through the PC reservoir individually, both of the excited states should be simultaneously resonant to the interacting photon, namely $\Delta_1 = \Delta_2$. Since there is no net energy transfer between atom and PC at the degenerate with $Q_T = 0$, the atom seems to have "allowed" transition between these two excited states through strong coupling with the PC reservoir. Therefore, the maximal quantum interference occurs near the degenerate. On the other hand, there is no energy exchange neither between two excited states with $Q_2 = 0$ nor between the atom and the PC with $Q_T = 0$ at the degenerate as the atom is equally in-phase prepared in two excited states ($\theta = \pi/4$), as shown in Fig. 4(b). In this case, the three-level atom is equivalent to a two-level system and there is one bounded DS existing in two-level atom with no quantum interference. Similar results are obtained when the atom is prepared equally out of phase in two excited states ($\theta = -\pi/4$).

We further define the coherence of two excited states as $\langle A_1^*(t) A_2(t) \rangle / \sqrt{\langle |A_1(t)|^2 \rangle \langle |A_2(t)|^2 \rangle}$ with $\langle \rangle$ being the time average to observe the phase relationship in the quantum interference. Figure 5 shows that the correlation of two excited states tends to be negative, which means the population of two excited states tends to be out of phase, except that two excited states are prepared in-phase or $\theta = \pi/4$ nearly degenerated. In general, the deeper the excited states are in the PBG, the larger the total population is, but the largest total population with complete coherence is near the degenerate when the excited states are prepared out of phase. The coherent coupling of two transitions in a PC causes no energy stored or propagation in the PC reservoir. Therefore, the total probability of two excited states in a V-type system can be controlled not only by the transition frequency detuning with respect to the photonic band edge of the PC but also the relative phase of initial states.

## VII. CONCLUSION

The dynamics of a V-type atom embedded in a PC is expressed by linear combination of four DSs in anisotropic model using fractional calculus. With different detuning of excited states with respect to the photonic band edge, the dynamic behavior can be classified into three regions: non-Markovian decay, damped quantum interference and quantum interference. When two excited states are above the band edge, the decay rates of the excited-state populations show non-Markovian with fast decay at the beginning but become slow after evolving a certain time. As only one excited state is within the band gap, the population reveals fast damped quantum interference initially, caused by the interference of decaying states, and finally reaches a stationary bound state in both excited states. In quantum interference regime, with both of the excited sates within the band gap, the quantum interference between two transitions is minimum near degenerate when two excited states of atom are initially either in-phase or out-of-phase equally prepared, but it is maximum near degenerate when the atom is initially prepared in one of the excited states.

In addition, the minimum energy exchange of the atom and PC occurs at degeneracy, i.e., $\Delta_1 \approx \Delta_2$ because it is equivalent to a two-level atom embedded into the PC in which there are no net energy exchange between the atom and PC after the amplitude of the decaying dressed states have approached to zero. The quantum interference caused by energy exchange of two transitions reveals that the forbidden-transition excited states can exchange energy via interaction with the PC reservoir even though there are no net energy exchange between the atom and PC, and the total population can be controlled not only by the amount of negative frequency detuning with respect to the band edge but also by the relative phase of initial states.

The correlation of two excited states tends to be negative. Therefore, there is the highest excited state population in a PC when two initial excited states are prepared out of phase.

## Acknowledgments

The authors would like to thank the National Science Council of the Republic of China for partial financial support under grants NSC99-2811-M-006-028, NSC 99-2112-M-006-017- MY3, NSC99-2221-E-009-095-MY3 and NSC99-2112-M-034 -002-MY3.

**Reference**
[1] E. Yablonovitch, Phys. Rev. Lett. **58**, 2059 (1987).
[2] S. John, and J. Wang, Phys. Rev. Lett. **64**, 2418 (1990).
[3] S. John, and J. Wang, Phys. Rev. B **43**, 12772 (1991).
[4] S. John, and T. Quang, Phys. Rev. A **50**, 1764 (1994).
[5] S. C. Cheng, J. N. Wu, T. J. Yang, and W. F. Hsieh, Phys. Rev. A **79**, 013801 (2009).
[6] J. N. Wu, C. H. Huang, S. C. Cheng, and W. F. Hsieh, Phys. Rev. A **81**, 023827 (2010).
[7] P. Lambropoulos, G. M. Nikolopoulos, T. R. Nielsen, and S. Bay, Rep. Prog. Phys. **63**, 455 (2000).
[8] S. C. Cheng, J. N. Wu, M. R. Tsai, and W. F. Hsieh, J. Phys.: Condens. Matter **21**, 015503 (2009).
[9] N. Vats, and S. John, Phys. Rev. A **58**, 4168 (1998).
[10] N. Vats, S. John, and K. Busch, Phys. Rev. A **65**, 043808 (2002).
[11] N. Vats, and T. Rudolph, J. Mod. Opt. **48**, 1495 (2001).
[12] S. Nishimura, N. Abrams, B. A. Lewis, L. I. Halaoui *et al.*, J. Am. Chem. Soc. **125**, 6306 (2003).
[13] Y. P. Yang, S. Y. Xie, H. Chen, S. Y. Zhu *et al.*, Opt. Commun. **182**, 349 (2000).
[14] Y. P. Yang, and S. Y. Zhu, Phys. Rev. A **61**, 043809 (2000).
[15] D. Yang, J. Wang, H. Z. Zhang, and J. B. Yao, J. Opt. B: Quantum Semicl. Opt. **40**, 1719 (2007).
[16] S. Bay, P. Lambropoulos, and K. Molmer, Phys. Rev. Lett. **79**, 2654 (1997).
[17] N. Foroozani, M. M. Golshan, and M. Mahjoei, Phys. Rev. A **76**, 015801 (2007).
[18] S. Y. Xie, and Y. P. Yang, Eur. Phys. J. D. **42**, 163 (2007).

[19] P. Meystre, and M. Sargent, *Elements of quantum optics* (Springer, New York, 2007).

[20] V. V. Kozlov, Y. Rostovtsev, and M. O. Scully, Phys. Rev. A **74**, 063829 (2006).

[21] Y. Yang, Z. X. Lin, S. Y. Zhu, H. Chen *et al.*, Phys. Lett. A **270**, 41 (2000).

[22] X. D. Sun, B. Zhang, and X. Q. Jiang, Opt. Commun. **281**, 5194 (2008).

[23] S. Y. Zhu, H. Chen, and H. Huang, Phys. Rev. Lett. **79**, 205 (1997).

[24] H. Nihei, and A. Okamoto, J. Mod. Opt. **51**, 1983 (2004).

[25] M. Barth, R. Schuster, A. Gruber, and F. Cichos, Phys. Rev. Lett. **96**, 243902 (2006).

[26] K. S. Miller, and B. Ross, *An introduction to the fractional calculus and fractional differential equations* (Wiley, New York, 1993).

[27] H. Z. Zhang, S. H. Tang, P. Dong, and J. He, J. Opt. B: Quantum Semicl. Opt. **4**, 300 (2002).

[28] M. Woldeyohannes, and S. John, J. Opt. B: Quantum Semicl. Opt. **5**, R43 (2003).

[29] C. K. Sun, F. Vallee, S. Keller, J. E. Bowers *et al.*, Appl. Phys. Lett. **70**, 2004 (1997).

[30] P. C. Ou, J. H. Lin, C. A. Chang, W. R. Liu *et al.*, J. Phys. D: Appl. Phys. **43**, 495103 (2010).

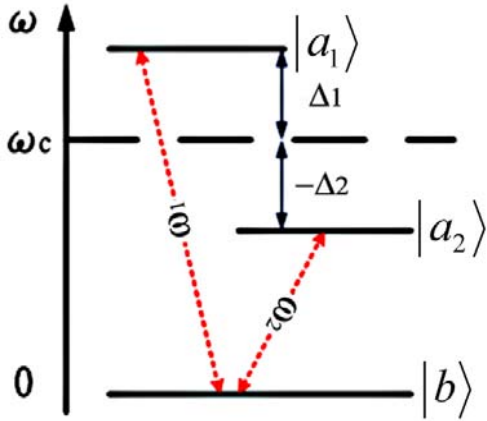

Fig. 1 Frequency diagram of a three-level atom placed in a photonic band gap structure. The two excited states, $|a_1\rangle$ and $|a_2\rangle$, detune from the photonic band edge ($\omega_c$) with $\Delta_1$ and $\Delta_2$.

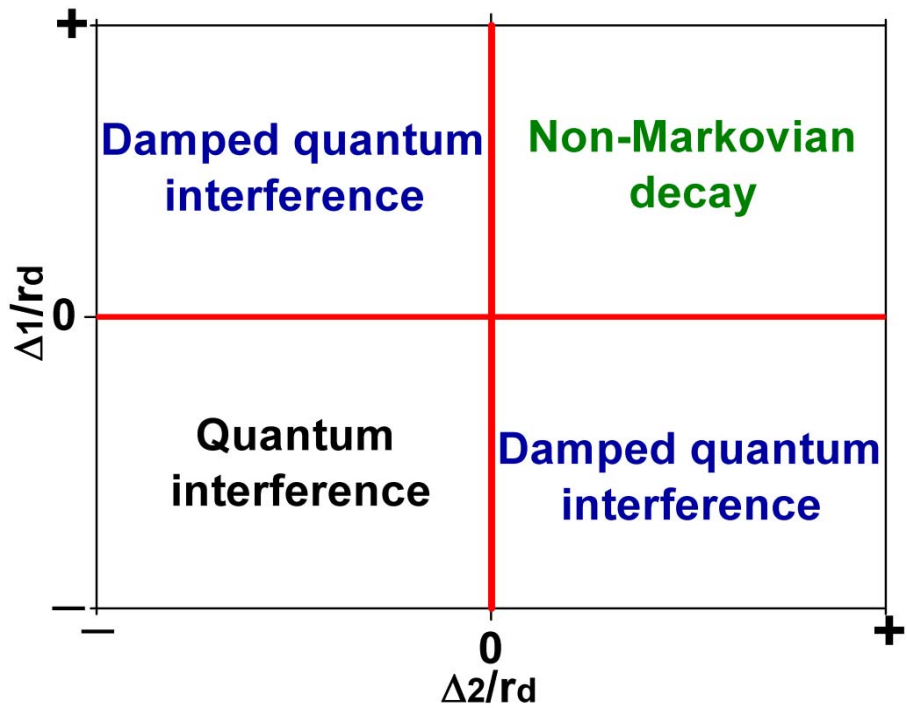

Fig. 2 Regions of SE dynamics in the anisotropic model.

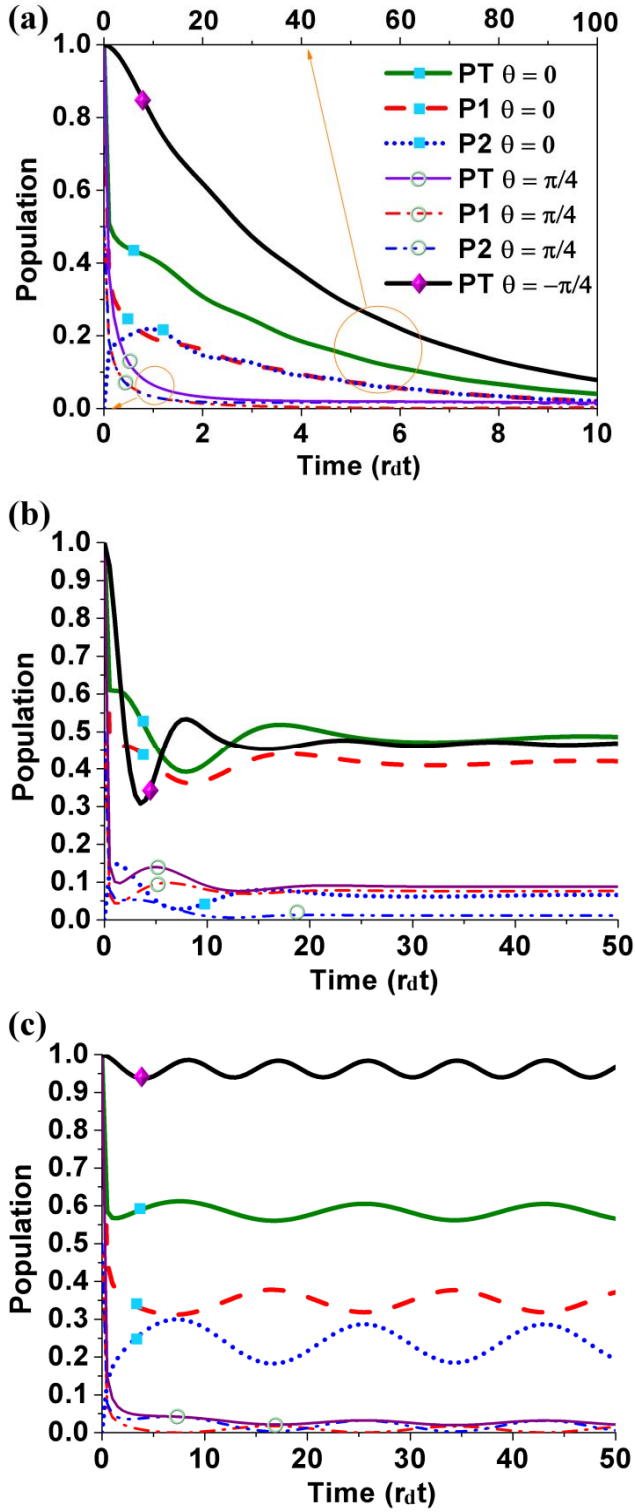

Fig. 3 Dynamics of SE in anisotropic PC systems. $P1 = |A_1|^2$ and $P2 = |A_2|^2$ are population of excited states 1 and 2 with (a) $\Delta_1/r_d = 0.5$ and $\Delta_2/r_d = 0.25$, (b) $\Delta_1/r_d = 0.5$ and $\Delta_2/r_d = -0.5$, and (c) $\Delta_1/r_d = -0.5$ and $\Delta_2/r_d = -0.25$. $PT$ equals to $P1+P2$.

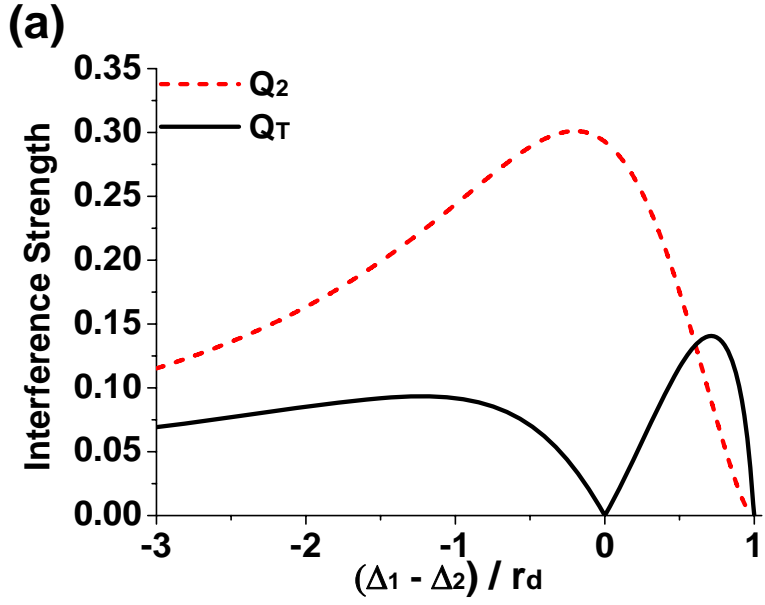

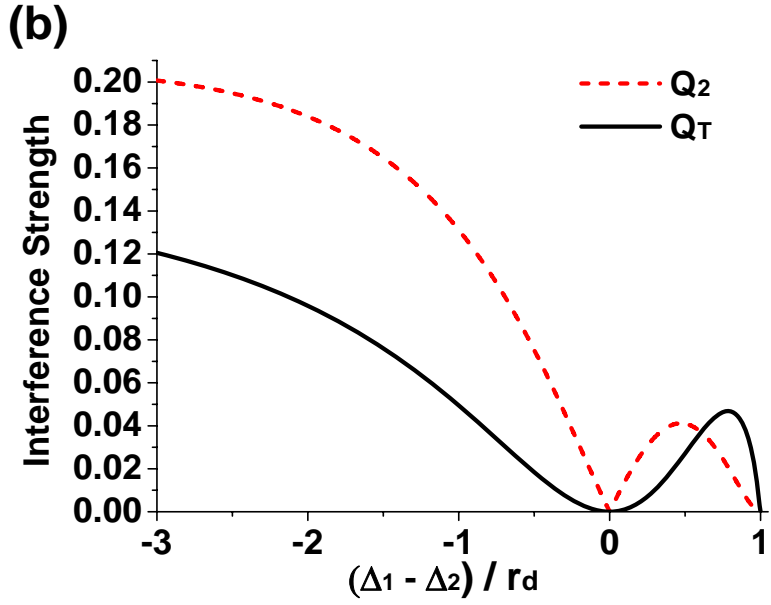

Fig. 4 Quantum interference strength of different initial condition with (a) $\theta = 0$, (b) $\theta = \pi/4$ and $\Delta_2 = -r_d$. The interference strength defines as the different of maximum and minimum population excited state two ($Q_2$) and total population ($Q_T$).

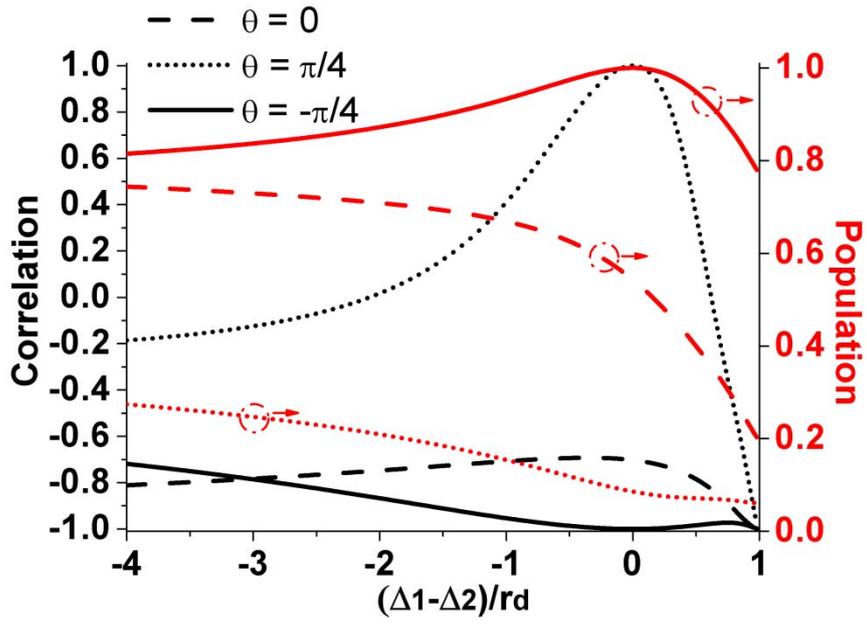

Fig. 5 Correlation and total population of different initial condition with $\Delta_2 = -r_d$.